\documentclass{article}
\usepackage{spconf}
\usepackage{graphicx}
\usepackage[cmex10]{amsmath}
\usepackage{amsfonts}
\usepackage{bm}
\usepackage{multirow}
\usepackage{color}
\title{Frequency-Undersampled Short-Time Fourier Transform\vspace{-3.3pt}}%
\name{Daichi \vspace{-3.3pt}Kitahara%
\thanks{This work was supported in part by JSPS Grants-in-Aid Grant Number for Early-Career Scientists JP19K20361 (e-mail: d-kita@fc.ritsumei.ac.jp).}}
\address{College of Information Science and Engineering, Ritsumeikan University, Shiga, \vspace{-3.4pt}Japan}%\vspace{-0.65pt}}

\begin{document}
\ninept
\maketitle
\begin{abstract}\vspace{-0.5pt}
The short-time Fourier transform (STFT) usually computes the same number of frequency components
as the frame length while overlapping adjacent time frames by more than half.
As a result, the number of components of a \textit{spectrogram matrix} becomes more than twice the signal length, and hence STFT is hardly used for signal compression. 
In addition, even if we modify the spectrogram into a desired one by spectrogram-based signal processing, it is re-changed during  the inversion as long as it is outside the range of STFT\@.
In this paper,~to~re-
duce the number of components of a spectrogram while maintaining the analytical ability, we propose the \textit{frequency-undersampled~STFT (FUSTFT)}, \hspace{-0.125pt}which \hspace{-0.125pt}computes \hspace{-0.125pt}only \hspace{-0.125pt}half \hspace{-0.125pt}the \hspace{-0.125pt}frequency \hspace{-0.125pt}components.
\hspace{-0.125pt}We also present the inversions with and without the \textit{periodic condition}, including their different properties.
In simple numerical examples~of audio signals, \vspace{-3pt}we confirm the validity of  FUSTFT and the inversions.%
\end{abstract}

\begin{keywords}%
Short-time Fourier transform, redundancy, spectrogram, inversions with and without periodicity, tridiagonal system.\vspace{-6.3pt}
\end{keywords}

\section{Introduction\vspace{-6.3pt}}
Time-frequency analysis is to capture the temporal variations of~frequency components of a target signal [1]--[18].
In audio signal processing, the~short-time~Fourier transform (STFT) [1]--[9] is the~most 
commonly used time-frequency analysis method since STFT inher-its the robustness, of the Fourier transform, against time shifts~\cite{Yatabe-Masuyama-Kusano-Oikawa19}.
The result of STFT is called a \textit{spectrogram}, and it is often expressed as a matrix. 
In addition to using spectrograms for signal analysis and feature extraction, we can also generate desired time-domain signals through \hspace{-0.11pt}modification \hspace{-0.11pt}of \hspace{-0.11pt}the \hspace{-0.11pt}spectrograms \hspace{-0.11pt}themselves \hspace{-0.11pt}[19]--[26]. \hspace{-0.11pt}This
paper is particularly aware of the latter usage of the spectrograms.

In most cases of the engineering field, STFT is used for~discrete-time signals, and a window function has a compact support.
In such a case, the support length of the discrete-time window function, called the \textit{window length}, is directly equal to  the length of each time frame, called the \textit{frame length}.
Typically, we calculate the same number of frequency components as the frame length in each time frame~by~using the fast Fourier transform (FFT).
We call this the~\textit{discrete STFT}\@.

We can also compute more frequency components, although that are \textit{linearly dependent}, in each time frame by padding zeros before FFT\@.
We call this the \textit{frequency-oversampled STFT (FOSTFT)}\@. Both the \hspace{-0.06pt}discrete \hspace{-0.06pt}STFT \hspace{-0.06pt}and \hspace{-0.06pt}FOSTFT \hspace{-0.06pt}are \hspace{-0.06pt}also \hspace{-0.06pt}called \hspace{-0.06pt}the \hspace{-0.06pt}\textit{windowed \hspace{-0.06pt}discrete Fourier transform (WDFT)} or the \textit{discrete Gabor transform (DGT)}, but in this paper we switch the names by focusing on the inequality between the frame length and the number of frequency components. 

The inversions, based on the \textit{Moore--Penrose pseudoinverse}, for the discrete STFT and FOSTFT can be easily computed by using~the so-called \textit{canonical dual window} \cite{Werther-Eldar-Subbanna-02}, \cite{Yang-08}, whose window length is the same as the frame length, after the inverse FFT (IFFT)\@.
These inversions for the discrete STFT and FOSTFT are called \textit{painless}~\cite{Daubechies-Grossmann-Meyer-86}.

From the facts that (i) human hearing is sensitive to block boundary artifacts and  (ii) a window function makes signal values small~at\pagebreak

\noindent both ends of each time frame, we usually overlap
adjacent frames by more \hspace{-0.018pt}than \hspace{-0.018pt}half \hspace{-0.018pt}in \hspace{-0.018pt}the \hspace{-0.018pt}computation \hspace{-0.018pt}of \hspace{-0.018pt}the \hspace{-0.018pt}discrete \hspace{-0.018pt}STFT \hspace{-0.018pt}and \hspace{-0.018pt}FOSTFT\@.

\noindent As a result, the number of components of a spectrogram matrix becomes more 
than twice the original signal length, and hence the spectrogram is hardly used for signal compression. 
Moreover, even if~we set components of a spectrogram to desired values by a spectrogram-based signal pressing technique such as [19]--[26],
 there is a risk that both magnitudes and phases would be greatly changed during the inversion unless the desired spectrogram belongs to the range of STFT\@.

As \textit{almost nonredundant} time-frequency analysis methods,\footnote{The redundancies of MDCT and DWT occur at the first and last frames.} the \textit{modified discrete cosine transform (MDCT)} \cite{Malvar-90}, that is used in coding formats for audio signals such as MP3 and AAC, and the \textit{discrete Wilson transform (DWT)} \cite{Daubechies-Jaffard-Journe-91}, \cite{Bolcskei-Feichtinger-Grochenig-Hlawatsch-96}, that is hardly used in an application because of a strict condition for a window function, are known.
The results of the discrete STFT and FOSTFT are complex-valued, while those of MDCT and DWT are real-valued and unsuitable for analysis of complex-valued signals.
Moreover, MDCT and DWT are sensitive to time shits differently from STFT\@.
As a complex version of MDCT, the \textit{modulated complex lapped transform (MCLT)} \cite{Malvar-99} is known \hspace{-0.1pt}but \hspace{-0.1pt}it \hspace{-0.1pt}is \hspace{-0.1pt}almost \hspace{-0.1pt}the \hspace{-0.1pt}same \hspace{-0.1pt}as \hspace{-0.1pt}the \hspace{-0.1pt}discrete \hspace{-0.1pt}STFT \hspace{-0.1pt}(see \hspace{-0.1pt}Footnote~\hspace{-0.1pt}8).

In this paper, to~suppress the redundancy of a spectrogram while maintaining the original analytical ability, we propose the \textit{frequency-undersampled STFT (FUSTFT)}, which calculates only half the frequency components of the discrete STFT in each time frame. 
From the fact that the energy of a target signal spreads along the~frequency axis by multiplying a  smooth window function, FUSTFT maintains the features of the original spectrogram of the discrete STFT despite  the undersampling.  
In fact, Stankovi\'{c}~has already proposed the special case of FUSTFT in \cite{Stankovic-16},  that is equivalent to Type-I FUSTFT in (11) with $\xi=\frac{L_{w}}{2}$.\footnote{Strictly \hspace{-0.17pt}speaking, \hspace{-0.17pt}sampling \hspace{-0.17pt}points \hspace{-0.17pt}of \hspace{-0.17pt}a \hspace{-0.17pt}window \hspace{-0.17pt}function \hspace{-0.17pt}are \hspace{-0.17pt}also~\hspace{-0.17pt}changed.}
Hence, this paper~is  the generalization of \cite{Stankovic-16}.

By using FUSTFT, we can easily obtain efficient spectrograms, including \hspace{-0.177pt}almost \hspace{-0.177pt}nonredundant \hspace{-0.177pt}ones,
\hspace{-0.177pt}while \hspace{-0.177pt}its \hspace{-0.177pt}inversion \hspace{-0.177pt}is \hspace{-0.177pt}not \hspace{-0.177pt}so~\hspace{-0.177pt}simple~differently from those for the discrete STFT and FOSTFT, i.e., its inversion process  changes dependently on the signal length \cite{Picot-Ferri-Herraez-Villanueva-18}. 
We realize the inversions with and without the \textit{periodic condition},~which is assumed in [6]--[8], by directly solving the least squares problems. 
In \cite{Picot-Ferri-Herraez-Villanueva-18} the general frequency-undersampling is considered~while this paper treats only the half frequency-undersampling and clarifies that both two different inversions can always be \vspace{-5.6pt}computed very quickly.

\section{Definitions of STFT and ISTFT in This Paper\vspace{-5.55pt}}
Let $\mathbb R$ and $\mathbb C$ be the sets of all real numbers and all complex numbers, respectively. 
The imaginary unit is denoted by $\imath\in\mathbb C$, i.e., $\imath^{2}=-1$. 
We write~vectors and matrices with boldface small and capital~letters, respectively. 
\hspace{-0.14pt}We \hspace{-0.14pt}express \hspace{-2.14pt}the \hspace{-0.14pt}transpose \hspace{-0.14pt}operator \hspace{-0.14pt}as \hspace{-0.14pt}$(\cdot)^{\mathrm{T}}\hspace{-1pt}$ \hspace{-0.14pt}and \hspace{-0.14pt}the~\hspace{-0.14pt}adjoint
operator as $(\cdot)^{\mathrm{H}}$.
We express the composition of mappings as $\hspace{1pt}\circ\hspace{1pt}$ and the inverse of a nonsingular matrix by $(\cdot)^{-1}$.
We express the $\ell_{2}$ norm of a vector as $\lVert \cdot \rVert_{\mathrm{2}}$ and the Frobenius norm
of a matrix as $\lVert \cdot \rVert_{\mathrm{F}}$. The floor and ceiling functions are denoted by $\lfloor \cdot \rfloor$ and $\lceil \cdot \rceil$, respectively.
For $a>0$, we define  $\mathrm{mod}_{a}:\mathbb R\to[0,a)$ by $\mathrm{mod}_{a}(b):=b-\bigl\lfloor \frac{b}{a}\bigr\rfloor a$.%

\subsection{Continuous-Time / Discrete-Time / Discrete STFT\vspace{-0.02pt}}
Let $x:\mathbb R\to \mathbb C$ be a real-valued or complex-valued continuous-time signal.
In this paper, with a real-valued window function $w:\mathbb R\hspace{-0.55pt}\to \hspace{-0.55pt}\mathbb R$, we define the \textit{continuous-time STFT} of \vspace{-1pt}$x(t)$ by\footnote{%
We use the sampling interval $T_{\mathrm{s}}$ in the definition of the continuous-time STFT
so that a discrete-time window function $w[\tau]$ in (5) will be symmetric.}%
\begin{equation}
X(f,t) =\int_{-\infty}^{\infty} x(\tau)\hspace{0.5pt}w(\tau-t)\hspace{0.5pt}e^{-\imath 2\pi f (\tau-t-\frac{T_{\mathrm{s}}}{2})}\,\mathrm{d}\tau\vspace{-1pt}
\end{equation}
and the \textit{discrete-time STFT} by\vspace{-2.15pt}\footnote{%
Note that we can define $X_{\mathrm{d}}(f,t)$ for all $t\in\mathbb R$ regardless of  $T_{\mathrm{s}}$ because the window function and the complex sinusoid are computable for any time~$t$.%
}%\vspace{-0.6pt}%
\begin{align}
X_{\mathrm{d}}(f,t)&= \sum_{\tau=-\infty}^{\infty} x(\tau\hspace{0.5pt} T_{\mathrm{s}})\hspace{0.5pt}w(\tau \hspace{0.5pt}T_{\mathrm{s}} -t)\hspace{0.5pt}e^{-\imath2\pi f (\tau \hspace{0.25pt}T_{\mathrm{s}}-t-\frac{T_{\mathrm{s}}}{2})}\hspace{-1.5pt}\\[-1.5pt]
&=\frac{1}{T_{\mathrm{s}}}\sum_{\kappa=-\infty}^{\infty}X(f-\kappa f_{\mathrm{s}},t)\mbox{,}\\[-15.5pt]\nonumber%\vspace{-1pt}
\end{align}
where $f\in\mathbb R$, $t \in\mathbb R$, $T_{\mathrm{s}} >0$ is the sampling interval of a discrete-time signal $x[\tau]:=x(\tau \hspace{0.5pt}T_{\mathrm{s}})$, and  $f_{\mathrm{s}}=\frac{1}{T_{\mathrm{s}}}$ is the sampling frequency.
From (3), $X_{\mathrm{d}}(f,t)$ is periodic  on $f$ with period $f_{\mathrm{s}}$, and hence we~can restrict $f$ to $f\in[-\frac{f_{\mathrm{s}}}{2},\frac{f_{\mathrm{s}}}{2})$ or $f\in[0,f_{\mathrm{s}})$ in the discrete-time STFT.

In what follows, let $L_{w}$ ($\geq2$) be an integer, and we suppose that the \hspace{-0.136pt}window \hspace{-0.136pt}function \hspace{-0.136pt}$w(t)$ \hspace{-0.136pt}has \hspace{-0.136pt}a \hspace{-0.136pt}compact \hspace{-0.136pt}support \hspace{-0.136pt}of \hspace{-0.136pt}length \hspace{-0.136pt}$L_{w}T_{\mathrm{s}}$,~\hspace{-0.136pt}i.e., $w(t)\neq0$ for almost all $t\in(0,L_{w}T_{\mathrm{s}})$ and $w(t)=0$~otherwise,~and $w(t)$ is a symmetric curve, i.e., $w(\frac{L_{w}T_{\mathrm{s}}}{2}-t)=w(\frac{L_{w}T_{\mathrm{s}}}{2}+t)$ for all $t\in\mathbb R$ and~$w'(t)\neq 0$ for almost all $t\in(0,L_{w}T_{\mathrm{s}})$. 
Under these assumptions,\footnote{%
The rectangular window is out of the discussion  because it is not a curve.%
} the continuous-time STFT in (1) is expressed as\vspace{-1.5pt}%
\begin{equation}
X(f,t) = \int_{0}^{L_{w}T_{\mathrm{s}}} x(\tau+t)\hspace{0.5pt}w(\tau)\hspace{0.5pt}e^{-\imath 2\pi f (\tau-\frac{T_{\mathrm{s}}}{2})}\,\mathrm{d}\tau\mbox{.}%\vspace{-1pt}
\end{equation}
In (2), we can define the discrete-time STFT for all~$t\in\mathbb R$, but there is almost no need to calculate $X_{\mathrm{d}}(f,t)$ at intervals shorter than $T_{\mathrm{s}}$.
Let $l$ be the time frame index.
With an integer frame shift $\xi$ ($\leq L_w$), we discretize the time $t$ of $X_{\mathrm{d}}(f,t)$ by \vspace{-2.15pt}$t=(l\xi-L_{w}+\xi-\tfrac{1}{2})\hspace{0.25pt} T_{\mathrm{s}}$,~i.e.,%
\begin{equation}
X_{\mathrm{d}}(f,\hspace{-0.25pt}(l\xi-L_{w}+\xi-\tfrac{1}{2})\hspace{0.25pt} T_{\mathrm{s}})
\hspace{-0.75pt}=\hspace{-2.51pt}\sum_{\tau=0}^{L_{w}-1}\hspace{-1.5pt}
x[\tau+l\xi-L_{w}+\xi]%x[\tau+l\xi]
\hspace{0.5pt} 
w[\tau]\hspace{0.5pt}e^{-\imath2\pi f \tau \hspace{0.25pt}T_{\mathrm{s}}\hspace{-0.25pt}}\mbox{,}\vspace{-1pt}
\end{equation}
where \vspace{-1pt}$f\hspace{-0.15pt}\in\hspace{-0.15pt}[0,f_{\mathrm{s}})$ and $w[\tau]\hspace{-0.15pt}:=\hspace{-0.15pt}w((\tau\hspace{1.15pt}+\hspace{1.15pt}\frac{1}{2})\hspace{0.25pt} T_{\mathrm{s}})\hspace{-0.15pt}\neq\hspace{-0.15pt} 0$. %
Next, let $k$~be~the 

\noindent frequency index, and discretize the frequency $f$ in (5) by $f=\tfrac{k}{L_{w}}f_{\mathrm{s}}$

\noindent since the maximum number of \textit{independent} frequency components computed in each time frame is $L_{w}$.
For a discrete-time~signal $\bm{x}:=(x[0],x[1],\ldots,x[L_x-1])^{\mathrm T}\in\mathbb C^{L_x}$ of length $L_{x}$ ($>L_{w}$), we define%
\begin{align}
&\mathrm{STFT}(\bm{x})[k,l]=X_{\mathrm{d}}(\tfrac{k}{L_{w}}f_{\mathrm{s}},(l\xi-L_{w}+\xi-\tfrac{1}{2}) \hspace{0.25pt}T_{\mathrm{s}})\nonumber\\[-1pt]
&\qquad=\sum_{\tau=0}^{L_{w}-1}x[\tau+l\xi-L_{w}+\xi] \hspace{0.5pt}w[\tau]\hspace{0.5pt}e^{-\imath2\pi\frac{ k}{L_{w}}\tau}\\[-15.5pt]\nonumber
\end{align}%
as the \textit{discrete STFT} in this paper,%
\footnote{%
For \textit{phase-aware} signal processing, it is shown in \vspace{-1.25pt}\cite{Yatabe-Masuyama-Kusano-Oikawa19} that another STFT
\begin{equation}
X(f,t)=\int_{0}^{L_{w}T_{\mathrm{s}}} x(\tau+t)\hspace{0.5pt}w(\tau)\hspace{0.5pt}e^{-\imath2\pi f (\tau+t)}\,\mathrm{d}\tau\vspace{-0.25pt}
\end{equation}
is better than (4) since complex spectrograms based on (7) will be lower~rank.
If we use two or more spectrograms with window functions of \textit{different}~$L_{w}$, it is better to change the support of $w(t)$ into \vspace{-2pt}$(-\frac{L_{w}T_{\mathrm{s}}}{2},\frac{L_{w}T_{\mathrm{s}}}{2})$ and compute
\begin{equation}
X(f,t)=\int_{-\frac{L_{w}T_{\mathrm{s}}}{2}}^{\frac{L_{w}T_{\mathrm{s}}}{2}} x(\tau+t)\hspace{0.5pt}w(\tau)\hspace{0.5pt}e^{-\imath2\pi f (\tau+t)}\,\mathrm{d}\tau\vspace{-1pt}
\end{equation}
instead of (7) since (8) aligns time frames of \textit{different lengths} at their~centers.%
} where $k=0,1,\ldots, L_{w}-1$~and%

\noindent $l\hspace{-0.2pt}=\hspace{-0.2pt}0,1,\ldots,\bigl\lceil\frac{L_{x} +L_{w} - 2\xi}{\xi}\bigr\rceil$. 
In \vspace{-0.75pt}(6), by assuming that $x(t)\hspace{-0.2pt}=\hspace{-0.2pt}0$ for~all $t\in(-\infty,-\frac{T_{\mathrm{s}}}{2}]\hspace{0.75pt}\cup\hspace{0.75pt}[(L_{x}\hspace{-0.5pt}-\frac{1}{2})\hspace{0.25pt}T_{\mathrm{s}},\infty)$, we padded $L_{w}\hspace{-0.5pt}-\xi$ zeros at the beginning of $\bm{x}$ and $\bigl\lceil\frac{L_{x} + L_{w} - \xi}{\xi}\bigr\rceil  \xi   - L_{x}$ zeros at the end.
The~discrete STFT in (6) is easily computed by FFT after multiplying the window function $w[\tau]$ and extracted time frame signals of length $L_{w}$. Unless $L_{w}$ \hspace{-0.05pt}is \hspace{-0.05pt}too \hspace{-0.05pt}large \hspace{-0.05pt}or \hspace{-0.05pt}$\xi$ \hspace{-0.05pt}is \hspace{-0.05pt}too \hspace{-0.05pt}small, \hspace{-0.05pt}a \hspace{-0.05pt}complex  %\\[-0.75pt]
spectrogram $\mathrm{STFT}(\bm{x})\hspace{-1.25pt}=$\\[-0.2pt]
$(\mathrm{STFT}(\bm{x})[k,l])\hspace{-0.35pt}\in\hspace{-0.35pt}\mathbb C^{L_{w}\times \left\lceil\frac{L_{x} + L_{w} - \xi}{\xi}\right\rceil}$ can be quickly obtained~\cite{Picot-Ferri-Herraez-Villanueva-18}.

In each time frame, we can also compute more frequency~components than $L_{w}$, that are \textit{linearly dependent}.
We call this transform the \textit{frequency-oversampled STFT (FOSTFT)}\@.
Specifically, let $N_{\mathrm{z}}$ be a \hspace{-0.1pt}positive \hspace{-0.1pt}integer, \hspace{-0.1pt}discretize \hspace{-0.1pt}$f$ \hspace{-0.1pt}in \hspace{-0.1pt}(5) \hspace{-0.1pt}by \hspace{-0.1pt}\vspace{-3.3pt}$f\hspace{-0.35pt}=\hspace{-0.35pt}\tfrac{k}{L_{w}+N_{\mathrm{z}}}f_{\mathrm{s}}$, \hspace{-0.1pt}and \hspace{-0.1pt}we~\hspace{-0.1pt}define
\begin{align}
&\mathrm{FOSTFT}(\bm{x})[k,l]=X_{\mathrm{d}}(\tfrac{k}{L_{w}+N_{\mathrm{z}}}f_{\mathrm{s}},(l\xi-L_{w}+\xi-\tfrac{1}{2})\hspace{0.25pt} T_{\mathrm{s}})\nonumber\\[-1pt]
&\qquad=\sum_{\tau=0}^{L_{w}-1}x[\tau+l\xi-L_{w}+\xi] \hspace{0.5pt}w[\tau]\hspace{0.5pt}e^{-\imath2\pi\frac{k}{L_{w}+N_{\mathrm{z}}}\tau}\mbox{,}\\[-18.2pt]\nonumber
\end{align}
where $k=0,1,\ldots, L_{w}+N_{\mathrm{z}}-1$ and $l=0,1,\ldots,\bigl\lceil\frac{L_{x} +L_{w} - 2\xi}{\xi}\bigr\rceil$.
FOSTFT in \vspace{-6.2pt}(9) is computed by padding $N_{\mathrm{z}}$ zeros right before~FFT\@.%

\subsection{Inversions for the Discrete STFT and FOSTFT\vspace{-2.2pt}}
The \hspace{-0.0539pt}discrete \hspace{-0.0539pt}STFT \hspace{-0.0539pt}in \hspace{-0.0539pt}(6) \hspace{-0.0539pt}is \hspace{-0.0539pt}a \hspace{-0.0539pt}\textit{linear \hspace{-0.0539pt}mapping} \hspace{-0.0539pt}and \hspace{-0.0539pt}we \hspace{-0.0539pt}express \hspace{-0.0539pt}its~\hspace{-0.0539pt}\textit{range}\\[-0.5pt]
as $\mathcal R\hspace{-0.7pt}:=\hspace{-0.7pt}\{\bm{X}\hspace{-0.7pt}\in\hspace{-0.7pt}\mathbb C^{L_{w}\times \left\lceil\frac{L_{x} + L_{w} - \xi}{\xi}\right\rceil}\,|\,\exists\bm{x}\hspace{-0.7pt}\in\hspace{-0.7pt}\mathbb C^{L_x}\, \bm{X}\hspace{-0.7pt} =\mathrm{STFT}(\bm{x})\}$. \hspace{-0.15pt}As 
long as $\xi<L_{w}$, the discrete STFT is \textit{redundant}, and there are innumerable linear mappings that recover, from a complex spectrogram $\bm{X}\in\mathcal R$, the corresponding signal $\bm{x}$ \cite{Werther-Eldar-Subbanna-02}. 
To recover the most~consistent \hspace{-0.1pt}signal \hspace{-0.1pt}$\bm{x}$ \hspace{-0.1pt}from \hspace{-0.1pt}\vspace{-1.5pt}$\bm{X}\hspace{-0.65pt}\not\in\hspace{-0.65pt}
\mathcal R$, 
 \hspace{-0.1pt}we \hspace{-0.1pt}define \hspace{-0.1pt}the \hspace{-0.1pt}\textit{inverse \hspace{-0.1pt}STFT \hspace{-0.1pt}(ISTFT)}~\hspace{0.1pt}by
\begin{equation}
\mathrm{ISTFT}(\bm{X})=\mathop{\mathrm{argmin}}_{\bm{x}\in\mathbb C^{L_x}} \,\lVert \bm{X} - \mathrm{STFT}(\bm{x})\rVert_{\mathrm{F}}^{2}\mbox{.}\vspace{-5.5pt}
\end{equation}

We express the discrete STFT as $\mathcal S:\mathbb C^{L_x}\hspace{-1pt}\to\mathbb C^{L_{w}\times \left\lceil\frac{L_{x} + L_{w} - \xi}{\xi}\right\rceil}$. 
Then, since ISTFT in (10) is the \textit{Moore--Penrose pseudoinverse} of~$\mathcal S$, 
we have $\mathrm{ISTFT}(\bm{X})\hspace{-0.75pt}=\hspace{-0.75pt}(\mathcal S^{\mathrm{H}}\hspace{0.5pt}\circ \hspace{1pt}\mathcal S)^{-1}\hspace{0.5pt}\circ \hspace{1pt}\mathcal S^{\mathrm{H}}(\bm{X})$. The~matrix $\mathcal S^{\mathrm{H}}\hspace{0.5pt}\circ\hspace{1pt} \mathcal S\in\mathbb R^{L_{w}\times L_{w}}$ is diagonal, and its diagonal components are periodic~with period $\xi$. 
Hence, ISTFT can be quickly computed by using IFFT and the pre-designed \textit{canonical dual window} \cite{Werther-Eldar-Subbanna-02}.
\hspace{-0.5pt}For FOSTFT in (9)~we can \vspace{-4.9pt}compute its inversion by using the same canonical dual~window.%

\section{Frequency-Undersampled STFT\vspace{-4.9pt}}
It is known that human hearing is sensitive to block boundary artifacts \cite{Malvar-90}.
Moreover, in each time frame, a non-rectangular window $w[\tau]$ makes signal values at both ends very small.
From these facts, we usually restrict the frame shift $\xi$ to $\xi \leq \frac{L_{w}}{2}$ in (6) and (9). However, in this usual case, the number of components of a~spectrogram matrix is more than twice the signal length $L_{x}$, which is not suitable 
 
\noindent for signal compression.
In addition, even if we obtain desired~spectrograms through spectrogram-based signal processing, their compo-nents will be changed by ISTFT unless they belong to the range $\mathcal R$.

In what follows, $L_{w}$ is a \textit{multiple of 4}.
For more efficient~time-frequency analysis, we propose the \textit{frequency-undersampled STFT (FUSTFT)}, that computes $\frac{L_{w}}{2}$ frequency components in each~frame.
We discretize $f$  in (5) by $f=\tfrac{2k}{L_{w}}f_{\mathrm{s}}$, and define \textit{Type-I FUSTFT}~as\vspace{-3.3pt}%
\begin{align}
&\mathrm{FUSTFT}_{\mathrm{I}}(\bm{x})[k,l]=X_{\mathrm{d}}(\tfrac{2k}{L_{w}}f_{\mathrm{s}},(l\xi-L_{w}+\xi-\tfrac{1}{2}) \hspace{0.25pt}T_{\mathrm{s}})\nonumber\\[-1pt]
&\qquad=\sum_{\tau=0}^{L_{w}-1}x[\tau+l\xi-L_{w}+\xi]\hspace{0.5pt}w[\tau] \hspace{0.5pt}e^{-\imath2\pi\frac{2k}{L_{w}}\tau}\mbox{.}\\[-17.1pt]\nonumber
\end{align}
Discretize $f$  in (5) by $f=\tfrac{2k+1}{L_{w}}f_{\mathrm{s}}$, and define \textit{Type-II FUSTFT}~\vspace{-3.3pt}as
\begin{align}\pagebreak
&\mathrm{FUSTFT}_{\mathrm{II}}(\bm{x})[k,l]=X_{\mathrm{d}}(\tfrac{2k+1}{L_{w}}f_{\mathrm{s}},(l\xi-L_{w}+\xi-\tfrac{1}{2}) \hspace{0.25pt}T_{\mathrm{s}})\nonumber\\[-1pt]
&\qquad=\sum_{\tau=0}^{L_{w}-1}x[\tau+l\xi-L_{w}+\xi]\hspace{0.5pt}w[\tau] \hspace{0.5pt}e^{-\imath2\pi\frac{2k+1}{L_{w}}\tau}\mbox{.}\\[-15.5pt]\nonumber
\end{align}

\noindent By using Type-I and Type-II alternately, define \vspace{-3.5pt}\textit{Type-III FUSTFT} as
\begin{align}
&\hspace{-9pt}\mathrm{FUSTFT}_{\mathrm{III}}(\bm{x})[k,l]\nonumber\\[-1pt]
&\hspace{-9pt}=\left\{%\hspace{-1pt}
\begin{aligned}
&\sum_{\tau=0}^{L_{w}-1}x[\tau+l\xi-L_{w}+\xi]\hspace{0.5pt}w[\tau] \hspace{0.5pt}e^{-\imath2\pi\frac{2k}{L_{w}}\tau}&&\mbox{if $l$ is even,}\\[-1pt]%[-1pt]%
&\sum_{\tau=0}^{L_{w}-1}x[\tau+l\xi-L_{w}+\xi]\hspace{0.5pt}w[\tau] \hspace{0.5pt}e^{-\imath2\pi\frac{2k+1}{L_{w}}\tau}&&\mbox{if $l$ is odd.}%\\
\end{aligned}\right.\hspace{-10pt}\\[-17.1pt]\nonumber
\end{align}%\vspace{-2.4pt}
From (11) to (13), $k\hspace{-0.72pt}=\hspace{-0.72pt}0,1,\ldots,\frac{L_{w}}{2}-1$, $l\hspace{-0.72pt}=\hspace{-0.72pt}0,1,\ldots,\bigl\lceil\frac{L_{x} +L_{w} - 2\xi}{\xi}\bigr\rceil$, and $\xi \leq \frac{L_{w}}{2}$.
As shown in Fig.~1,\footnote{In Fig.~1, \vspace{-0.5pt}$f_{\mathrm{s}}=44{,}100$ [Hz], $L_{w}=512$, $\xi =\frac{L_w}{2}=256$, and we used the normalized sine window $w[\tau]:=\frac{1}{\sqrt{L_w}}\sin(\frac{1}{L_w}(\tau+\frac{1}{2})\pi)$ for both (6) and (12). 
Since a sound $\bm{x}$ is real-valued, one in each complex conjugate pair, e.g.,~%
$\mathrm{FUSTFT}_{\mathrm{II}}(\bm{x})[k,l]$ and $\mathrm{FUSTFT}_{\mathrm{II}}(\bm{x})[\frac{L_w}{2}-k-1,l]$, was omitted.%
}
the number of frequency bins of FUSTFT is half of that of the discrete STFT\@.\footnote{Combining FOSTFT of \vspace{-2.6pt}$N_{\mathrm{z}}\hspace{-0.25pt}=\hspace{-0.25pt}L_{w}$ and Type-II FUSTFT, we can~\textit{redefine}%
\begin{equation}
\mathrm{STFT}(\bm{x})[k,l]
%\\
=\sum_{\tau=0}^{L_{w}-1}x[\tau+l\xi-L_{w}+\xi]\hspace{0.5pt}w[\tau] \hspace{0.5pt}e^{-\imath2\pi\frac{2k+1}{2L_{w}}\tau}%\vspace{-0.5pt}
\end{equation}
as the discrete STFT\@.
If we express the transform in (14) as a linear mapping $\mathcal S$, its inversion is easily computed since $\mathcal S^{\mathrm{H}}\circ \mathcal S$ is the same as (6). (14) with\\[-1pt]
$\xi=\frac{L_{w}}{2}$ and MCLT~\cite{Malvar-99} have the \textit{same magnitudes} and differ only in phases.}
Since the energy~of~$\bm{x}$ spreads along the frequency axis by multiplying a % smooth 
window function, FUSTFT maintains the characteristics of the standard~spectrogram~of the discrete STFT despite  \vspace{-5pt}undersampling of frequency components.

\begin{figure}
%\begin{minipage}{0.5\hsize}
\begin{minipage}{0.235\textwidth}
\centering
          \includegraphics[width=\hsize]{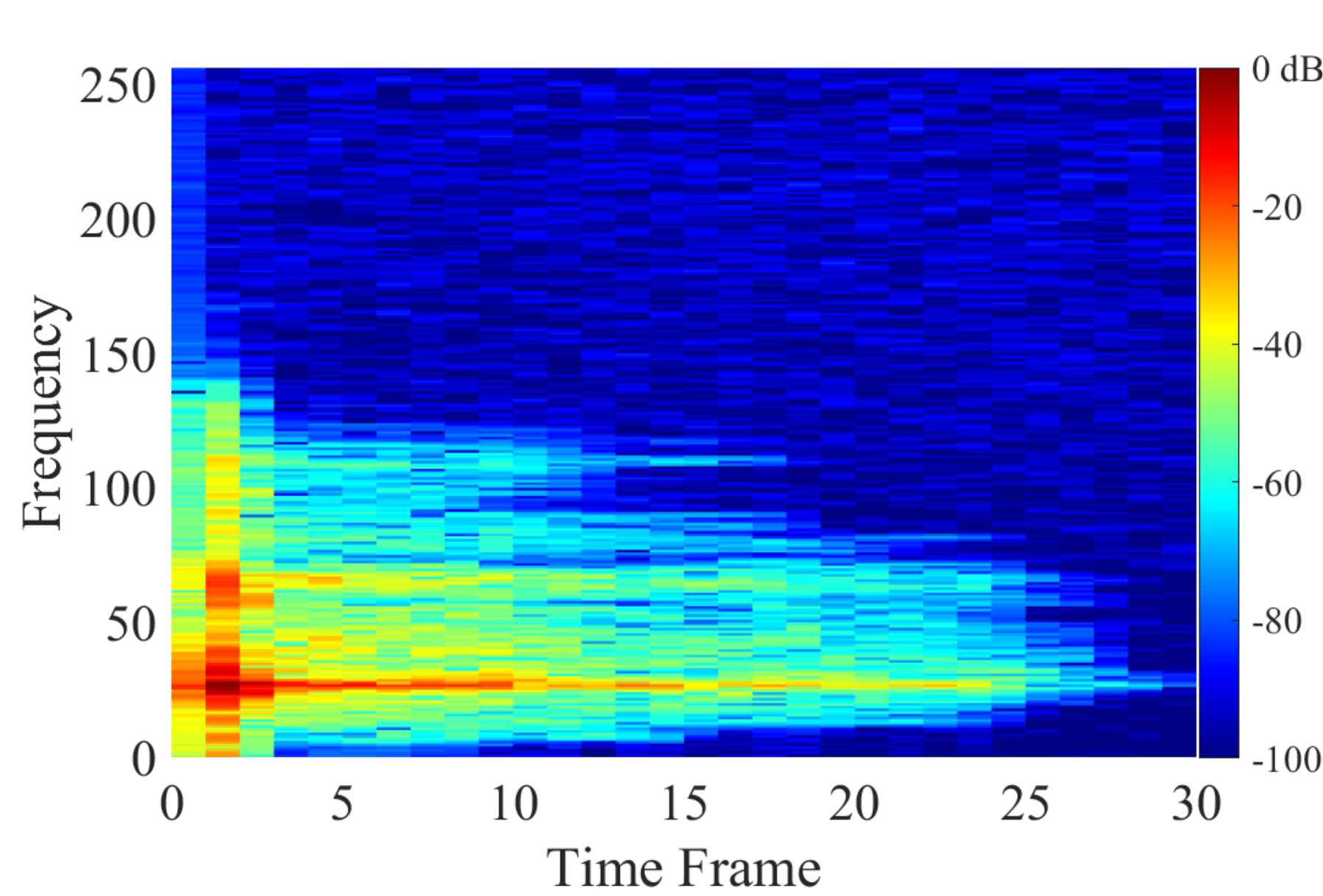}\\%[-2pt]
           (a) Discrete STFT in (6)%
          \end{minipage}\ \ %
% 1-2
      \begin{minipage}{0.235\textwidth}
      \centering
          \includegraphics[width=\hsize]{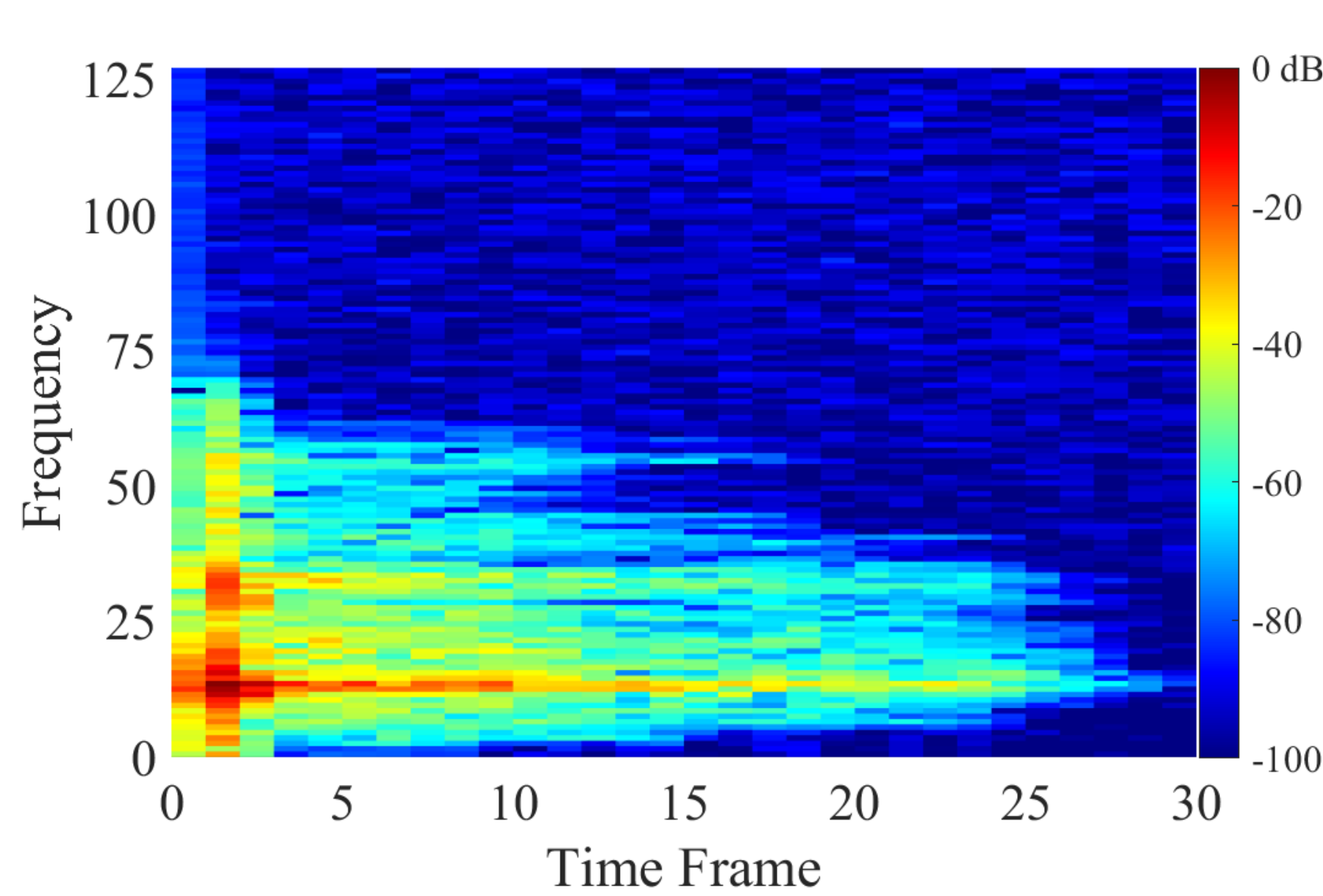}\\%[-2pt]
           (b) Type-II FUSTFT in (12)%
          \end{minipage}\\[-8pt]%[-5pt]%
          \caption{Power spectrograms of a clicking sound of $0.2$ seconds \cite{mat}.}%\vspace{-1pt}
          \vspace{-7pt}%
\end{figure}

\section{Two Different Inversions for FUSTFT \vspace{-4.2pt}}
\subsection{Inversion Based on the Standard Pseudoinverse\vspace{-2.2pt}}
Differently from cases of the discrete STFT and FOSTFT, the inversion for FUSTFT is not simple, i.e., the canonical dual window~\textit{does not exist} and the computation process changes dependently on~$L_{x}$.  
In this paper, we realize the inversion by solving the problem~similar\\[-0.75pt]
to (10). Let $\mathcal S:\mathbb C^{L_x}\hspace{-1pt}\to\mathbb C^{\frac{L_{w}}{2}\times \left\lceil\frac{L_{x} + L_{w} - \xi}{\xi}\right\rceil}$ be one of the linear~mappings (11), (12), and (13).
Then the inversion based on the pseudo-inverse is expressed as $(\mathcal S^{\mathrm{H}}\circ \mathcal S)^{-1}\circ \mathcal S^{\mathrm{H}}$. 
In FUSTFT cases, $\mathcal S^{\mathrm{H}}\circ \mathcal S$~is\vspace{-4pt}%
\setcounter{MaxMatrixCols}{13}
\arraycolsep=2pt
\begin{equation}
\mbox{\footnotesize$
\hspace{-2.4pt}
\begin{bmatrix}
\hspace{4pt}a_{0}&&&\hspace{2pt}b_{0}&&&&&\\[-2pt]
&\hspace{3pt}a_{1}&&&\hspace{-5pt}b_{1}&&&&&\\[-7pt]
&&\hspace{3pt}\ddots&&&\hspace{-4pt}\ddots&&&&&\\[-3pt]
\hspace{4pt}b_{0}&&&\hspace{2pt}a_{\hspace{-1pt}\frac{L_{w}}{2}}&&&b_{\hspace{-1pt}\frac{L_{w}}{2}}\\[-6pt]
&\hspace{3pt}b_{1}&&&\hspace{-5pt}a_{\hspace{-1pt}\frac{L_{w}}{2}\hspace{-1pt}+\hspace{-1pt}1}&&&\hspace{-5pt}\ddots\\[-6pt]
&&\hspace{3pt}\ddots&&&\hspace{-4pt}\ddots&&&\hspace{-4pt}\ddots&&&\\[-6pt]
&&&\hspace{2pt}b_{\hspace{-1pt}\frac{L_{w}}{2}}&&&\ddots&&&\hspace{11pt}\ddots\\[-6pt]
&&&&\hspace{-5pt}\ddots&&&\hspace{-5pt}a_{L_{x}\hspace{-1pt}-\hspace{-1pt}\frac{L_{w}}{2}\hspace{-1pt}-\hspace{-1pt}1}&&&b_{L_{x}\hspace{-1pt}-\hspace{-1pt}\frac{L_{w}}{2}\hspace{-1pt}-\hspace{-1pt}1\hspace{-1.5pt}}\\[-6pt]
&&&&&\hspace{-4pt}\ddots&&&\hspace{-4pt}\ddots\\[-4pt]
&&&&&&\ddots&&&\hspace{11pt}\ddots\\[-3pt]
&&&&&&&\hspace{-5pt}b_{L_{x}\hspace{-1pt}-\hspace{-1pt}\frac{L_{w}}{2}\hspace{-1pt}-\hspace{-1pt}1}&&&a_{L_{x}\hspace{-1pt}-\hspace{-1pt}1\hspace{-1.5pt}}
\end{bmatrix}$}\hspace{-0.1pt}\mbox{.}\vspace{-3pt}\pagebreak
\end{equation}

\noindent In (15),\footnote{Let $\bm{P}_{l}\hspace{-1.25pt}\in\hspace{-1.25pt}\mathbb R^{L_w\hspace{-0.5pt}\times \hspace{-0.25pt}L_x\hspace{-1pt}}_{+}$, $\bm{W}\hspace{-1.15pt}=\hspace{-1.15pt}\mathrm{diag}(w[\tau])\hspace{-1.25pt}\in\hspace{-1.25pt}\mathbb R^{L_w\hspace{-0.5pt}\times\hspace{-0.25pt} L_w\hspace{-1pt}}_{+}$, and $\bm{F}_{\mathrm{u}}\hspace{-1.25pt}\in\hspace{-1.25pt}\mathbb C^{\hspace{-0.25pt}\frac{L_w}{2}\hspace{-0.5pt}\times\hspace{-0.25pt} L_w}$ be
the $l$th frame extraction matrix, window matrix, and Type-I undersampled DFT matrix.
\hspace{-0.4pt}We have $\mathcal S^{\mathrm{H}}\hspace{0.3pt}\circ\hspace{0.8pt} \mathcal S\hspace{-1.1pt}=\hspace{-1.3pt}\sum_{l}\hspace{-1pt}\bm{P}^{\mathrm{T}}_{l}\hspace{-0.2pt}\bm{W}\hspace{-0.2pt}\bm{F}_{\mathrm{u}}^{\mathrm{H}}\hspace{-0.2pt}\bm{F}_{\mathrm{u}}\hspace{-0.2pt}\bm{W}\hspace{-0.2pt}\bm{P}_{l}\hspace{-0.2pt}$ for Type-I~FUSTFT.} diagonal components are\vspace{-3.4pt}%
\begin{equation}
a_{i}=\frac{L_{w}}{2}\sum_{l=0}^{\left\lceil\frac{L_{w}}{\xi}\right\rceil-1}w^{2}[m_{i}+l\xi]\vspace{-3.4pt}
\end{equation}
for all the three types, where $m_{i}\hspace{-0.25pt}=\hspace{-0.25pt}\mathrm{mod}_{\xi}(\hspace{0.25pt}i+L_{w}\hspace{-0.25pt})$ and $w[\tau]\hspace{-0.25pt}=\hspace{-0.25pt}0$~for $\tau\geq L_{w}$.
For Type-I FUSTFT, nonzero nondiagonal components are\vspace{-3.4pt}%
\begin{equation}
b_{i}=\frac{L_{w}}{2}\sum_{l=0}^{\left\lceil\frac{L_{w}}{2\xi}\right\rceil-1}w[m_{i}+l\xi]\hspace{0.5pt}w[m_{i}+l\xi+\tfrac{L_{w}}{2}]\mbox{.}\vspace{-3.4pt}%\vspace{-1pt}
\end{equation}
For Type-II, each $b_{i}$ is equal to (17) multiplied by $-1$.
For \vspace{-3.4pt}Type-III,%\vspace{-8pt}%
\begin{equation}%\pagebreak
b_{i}=\frac{L_{w}}{2}
\sum_{l=0}^{\left\lceil\frac{L_{w}}{2\xi}\right\rceil-1}
(-1)^{\left\lfloor\frac{i+L_{w}-\xi}{\xi}\right\rfloor+l}\hspace{0.5pt}w[m_{i}+l\xi]\hspace{0.5pt}w[m_{i}+l\xi+\tfrac{L_{w}}{2}]\mbox{.}\vspace{-1pt}%\vspace{-1pt}
\end{equation}%
$a_{i}$ in (16) and $b_{i}$ in (17) are periodic with period $\xi$ while $b_{i}$ in (18)~is periodic \hspace{-0.1pt}with \hspace{-0.1pt}period \hspace{-0.1pt}$2\xi$. 
\hspace{-0.5pt}We \hspace{-0.1pt}only \hspace{-0.1pt}have \hspace{-0.1pt}to \hspace{-0.1pt}compute \hspace{-0.1pt}them \hspace{-0.1pt}for \hspace{-0.1pt}one~\hspace{-0.1pt}cycle.\vspace{-1pt}%

For a given complex spectrogram $\bm{X}\in\mathbb C^{\frac{L_{w}}{2}\times \left\lceil\frac{L_{x} + L_{w} - \xi}{\xi}\right\rceil}$, define $\bm{y}=:(y[0],y[1],\ldots,y[L_x-1])^{\mathrm T}:=\mathcal S^{\mathrm{H}}(\bm{X})$.
Then, the~unique solution $\bm{x}$ to a linear system $(\mathcal S^{\mathrm{H}}\circ\mathcal  S)\hspace{1pt}\bm{x} = \bm{y}$ is the inversion result~of $\bm{X}$\@.
This linear system is decomposed into \vspace{-3.4pt}$\frac{L_{w}}{2}$ independent systems
\arraycolsep=1pt
\begin{equation}
\mbox{\small$
\hspace{-1.9pt}\begin{bmatrix}
a^{\langle i \rangle}_{0}\hspace{-1pt}&b^{\langle i \rangle}_{0}\hspace{-1pt}&&&\\[0.5pt]
b^{\langle i \rangle}_{0}\hspace{-1pt}&a^{\langle i \rangle}_{1}\hspace{-1pt}&\hspace{-3pt}b^{\langle i \rangle}_{1}\\[-4pt]
&\hspace{-1pt}\ddots&\hspace{3pt}\ddots&\hspace{-4pt}\ddots\\[-1pt]
&&\!\!\hspace{0.5pt}b^{\langle i \rangle}_{n_i-3}\hspace{-0.5pt}&a^{\langle i \rangle}_{n_i-2}\hspace{-1.5pt}&b^{\langle i \rangle}_{n_i-2}\\[1.5pt]
&&&b^{\langle i \rangle}_{n_i-2}\hspace{-1.5pt}&a^{\langle i \rangle}_{n_i-1}
\end{bmatrix}
\hspace{-5pt}\begin{bmatrix}
x[i]\\[2pt]
x[i\!+\hspace{-1pt}\!\tfrac{L_{w}}{2}]\\[-3pt]
\vdots\\%[-1pt]
x[i\!+\hspace{-1pt}\!\tfrac{(n_i-2)L_{w}}{2}]\\[2pt]
x[i\!+\hspace{-1pt}\!\tfrac{(n_i-1)L_{w}}{2}]
\end{bmatrix}
\hspace{-4.1pt}=\hspace{-4.1pt}
\begin{bmatrix}
y[i]\\[2pt]
y[i\!+\hspace{-1pt}\!\tfrac{L_{w}}{2}]\\[-3pt]
\vdots\\%[-1pt]
y[i\!+\hspace{-1pt}\!\tfrac{(n_i-2)L_{w}}{2}]\\[2pt]
y[i\!+\hspace{-1pt}\!\tfrac{(n_i-1)L_{w}}{2}]
\end{bmatrix}$}\hspace{-2pt}\vspace{-1pt}
\end{equation}
($i=0,1,\ldots,\frac{L_{w}}{2}-1$), where $n_{i}=\bigl\lceil\frac{2(L_{x}-i)}{L_{w}}\bigr\rceil$,
 $a^{\langle i \rangle}_{j}\hspace{-1pt}:=a_{i+j\frac{L_{w}}{2}}$~and\\[-2.25pt]
  $b^{\langle i \rangle}_{j}\hspace{-1pt}:=b_{i+j\frac{L_{w}}{2}}$.
Since the matrices in the left side of (19) are \vspace{-1pt}\textit{tridiagonal matrices}, their LU decompositions can be computed in $\mathcal O(n)$ \cite{Thomas-49}, and the inversion result $\bm{x}$ is also obtained from $\bm{y}$ in $\mathcal O(n)$.\footnote{The solver for tridiagonal systems is called the Thomas algorithm.}\vspace{0.75pt} %()

In particular, when $\mathrm{mod}_{\xi}(\frac{L_{w}}{2})=0$ for Type-I and Type-II,~\vspace{-0.5pt}or $\mathrm{mod}_{2\xi}(\frac{L_{w}}{2})\hspace{-0.22pt}=\hspace{-0.22pt}0$ for Type-III, we have $a^{\langle i \rangle}_{0}\hspace{-1.22pt}=\hspace{-0.22pt}a^{\langle i \rangle}_{1}\hspace{-1.22pt}=\hspace{-0.22pt}\cdots\hspace{-0.22pt}=\hspace{-0.22pt}a^{\langle i \rangle}_{n_{i}-1}\hspace{-1.22pt}=a_{i}$ and $b^{\langle i \rangle}_{0}\hspace{-1pt}=b^{\langle i \rangle}_{1}\hspace{-1pt}=\cdots=b^{\langle i \rangle}_{n_{i}-2}\hspace{-1pt}=b_{i}$, and hence the matrices in~the left side of (19) are \textit{tridiagonal Toeplitz matrices}.
In such cases, the eigenvalues are \vspace{-0.75pt}$\lambda^{\langle i \rangle}_{q}\hspace{-1pt}=a_{i}+2b_{i}\cos(\frac{q\pi}{n_{i}+1})>a_{i}-2|b_{i}|>0$ and the eigenvectors are
$\bm{u}_{q}=(\sin(\frac{q\pi}{n_{i}+1}), \sin(\frac{2q\pi}{n_{i}+1}),\ldots,\sin(\frac{n_{i}q\pi}{n_{i}+1}))^{\mathrm T}\in\mathbb R^{n_{i}}$ ($q=1,2,\ldots,n_{i}$) \cite{Noschese-Pasquini-Reichel-13}.
Therefore, the inversion result $\bm{x}$ is also obtained by using the discrete sine transform (DST) of Type-I \cite{Martucci-94}.\vspace{0.75pt}%

When \vspace{-0.75pt}$\mathrm{mod}_{\xi}(\frac{L_{w}}{2})=0$ and $\mathrm{mod}_{2\xi}(\frac{L_{w}}{2})\neq0$ for Type-III, we have $b^{\langle i \rangle}_{j}\hspace{-1pt}=(-1)^{j}b_{i}$ for all $i$ and $j$.
We can also compute the inversion result $\bm{x}$ by using Type-I DST, even in this case,  with appropriate sign reversal process \vspace{-6.2pt}(see the actual program in \cite{program} for more~detail).%

\subsection{Inversion Based on the Pseudoinverse with the Periodicity\vspace{-2.2pt}}
For Type-I and Type-II, let $p$ be the minimum nonnegative integer~s.t. $\mathrm{mod}_{\frac{L_w}{2}}(\bigl\lceil\frac{L_{x} +L_{w} - \xi}{\xi}\bigr\rceil \hspace{0.25pt}\xi+ p\hspace{1pt}\xi)=0$. 
For Type-III, $p$ must also satisfy\\[-0.75pt]%
 $\mathrm{mod}_{2}(\bigl\lceil\frac{L_{x} +L_{w} - \xi}{\xi}\bigr\rceil\hspace{-0.25pt}+ p)=0$.
\vspace{-0.5pt}We define $L_{\mathrm{p}}\hspace{-1pt}:=\bigl\lceil\frac{L_{x} +L_{w} - \xi}{\xi}\bigr\rceil \hspace{1pt}\xi+ p\hspace{1pt}\xi $ and $\bm{x}_{\mathrm{p}}:=(x[0],x[1],\ldots,x[L_x-1],x[L_x],\ldots,x[L_{\mathrm{p}}-1])^{\mathrm T}:=(\bm{x}^{\mathrm{T}}\hspace{-0.5pt},\hspace{-0.5pt}\bm{0}^{\mathrm{T}}_{L_{\mathrm{p}}-L_x})^{\mathrm{T}}\hspace{-1.5pt}\in\hspace{-0.5pt}\mathbb C^{L_{\mathrm{p}}}$.
Let ${\mathcal S}_{\mathrm{p}}\hspace{-1pt}:\mathbb C^{L_{\mathrm{p}}}\hspace{-1.5pt}\to\hspace{-0.5pt}\mathbb C^{\frac{L_{w}}{2}\times \left(\left\lceil\frac{L_{x} + L_{w} - \xi}{\xi}\right\rceil+p\right)}\hspace{-0.5pt}$~be a linear mapping, that computes one of (11), (12), and (13) for
\vspace{1pt}$l=0,1,\ldots,\bigl\lceil\frac{L_{x} +L_{w} - 2\xi}{\xi}\bigr\rceil \hspace{-0.75pt}+\hspace{-0.5pt}p\hspace{0.5pt}$ while assuming the \textit{periodic condition}, i.e., $x[\tau]\hspace{-0.5pt}=\hspace{-0.5pt}x[L_{\mathrm{p}}\hspace{0.75pt}+\hspace{0.75pt}\tau]$ for $\tau\hspace{-0.5pt}<\hspace{-0.5pt}0$, according to the convention [6]--[8].\pagebreak%

\mbox{}\\[-12.8pt]
\indent For a given complex spectrogram \vspace{-0.5pt}$\bm{X}\in\mathbb C^{\frac{L_{w}}{2}\times \left\lceil\frac{L_{x} + L_{w} - \xi}{\xi}\right\rceil}$,~we define \vspace{-0.5pt}$\bm{X}_{\mathrm{p}}\hspace{-1pt}:=\hspace{-0.3pt}[\bm{X},\bm{O}_{\frac{L_{w}}{2}\times p\hspace{0.5pt}}]\hspace{-0.5pt}\in\hspace{-0.5pt}\mathbb C^{\frac{L_{w}}{2}\times\left(\left\lceil\frac{L_{x} +L_{w} - \xi}{\xi}\right\rceil +p\right)}\hspace{-0.5pt}$ and compute 
 the \hspace{-0.1pt}inversion \hspace{-0.1pt}result \hspace{-0.1pt}$\bm{x}_{_{\mathrm{p}}}\hspace{-2pt}=\hspace{-0.5pt}(\mathcal S^{\mathrm{H}}_{\mathrm{p}}\hspace{0.1pt}\circ \hspace{0.5pt}\mathcal S_{\mathrm{p}})^{-1}\hspace{0.1pt}\circ \hspace{0.5pt}\mathcal S^{\mathrm{H}}_{\mathrm{p}}(\bm{X}_{\mathrm{p}})$
  \hspace{-0.1pt}of \hspace{-0.1pt}$\bm{X}_{\mathrm{p}}\hspace{-1pt}$ \hspace{-0.1pt}under \hspace{-0.1pt}the~\hspace{-0.1pt}pe-riodic condition.
Then, by extracting the first $L_x$ components~of $\bm{x}_{\mathrm{p}}$,

\noindent we obtain the final inversion result $\bm{x}$ of $\bm{X}$.
Therefore, define
$\bm{y}_{\mathrm{p}}\hspace{-1pt}:=(y[0],y[1],\ldots,y[L_{\mathrm{p}}-1])^{\mathrm T}:=\mathcal S^{\mathrm{H}}_{\mathrm{p}}(\bm{X}_{\mathrm{p}})$, and we only have to~compute the unique solution $\bm{x}_{\mathrm{p}}\hspace{-0.5pt}$ to  a linear system $(\mathcal S^{\mathrm{H}}_{\mathrm{p}}\hspace{-0.5pt}\circ \mathcal S_{\mathrm{p}})\hspace{1pt}\bm{x}_{\mathrm{p}}\hspace{-1pt} = \bm{y}_{\mathrm{p}}$.
This is decomposed into \vspace{-2.4pt}$\frac{L_{w}}{2}$ independent systems of the same size\footnote{Unless we expand $\bm{X}$ to $\bm{X}_{\mathrm{p}}$ by concatenating the appropriate zero matrix, the linear system $(\mathcal S^{\mathrm{H}}_{\mathrm{p}}\hspace{-0.5pt}\circ \mathcal S_{\mathrm{p}})\hspace{1pt}\bm{x}_{\mathrm{p}}\hspace{-1pt} = \bm{y}_{\mathrm{p}}$ cannot be decomposed into (20).}%
\arraycolsep=1pt%
\begin{equation}
\mbox{\small$
\hspace{-1.9pt}\begin{bmatrix}
a^{\langle i \rangle}_{0}\hspace{-1pt}&b^{\langle i \rangle}_{0}\hspace{-1pt}&&&b^{\langle i \rangle}_{n_{i}-1}\hspace{-1pt}\\
b^{\langle i \rangle}_{0}\hspace{-1pt}&a^{\langle i \rangle}_{1}\hspace{-1pt}&\hspace{-3pt}b^{\langle i \rangle}_{1}\\[-4pt]
&\hspace{-1pt}\ddots&\hspace{3pt}\ddots&\hspace{-4pt}\ddots\\[-1pt]
&&\!\!\hspace{0.5pt}b^{\langle i \rangle}_{n_i-3}\hspace{-0.5pt}&a^{\langle i \rangle}_{n_i-2}\hspace{-1.5pt}&b^{\langle i \rangle}_{n_i-2}\\[2pt]
b^{\langle i \rangle}_{n_i-1}\hspace{-8.50906pt}&&&b^{\langle i \rangle}_{n_i-2}\hspace{-1.5pt}&a^{\langle i \rangle}_{n_i-1}
\end{bmatrix}
\hspace{-5pt}\begin{bmatrix}
x[i]\\[2pt]
x[i\!+\hspace{-1pt}\!\tfrac{L_{w}}{2}]\\[-3pt]
\vdots\\%[-1pt]
x[i\!+\hspace{-1pt}\!\tfrac{(n_i-2)L_{w}}{2}]\\[2pt]
x[i\!+\hspace{-1pt}\!\tfrac{(n_i-1)L_{w}}{2}]
\end{bmatrix}
\hspace{-4.1pt}=\hspace{-4.1pt}
\begin{bmatrix}
y[i]\\[2pt]
y[i\!+\hspace{-1pt}\!\tfrac{L_{w}}{2}]\\[-3pt]
\vdots\\%[-1pt]
y[i\!+\hspace{-1pt}\!\tfrac{(n_i-2)L_{w}}{2}]\\[1.5pt]
y[i\!+\hspace{-1pt}\!\tfrac{(n_i-1)L_{w}}{2}]
\end{bmatrix}$}\hspace{-2pt}\vspace{1pt}
\end{equation}
($i=0,1,\ldots,\frac{L_{w}}{2}-1$), \vspace{-2.5pt}where $n_{0}=n_{1}=\cdots=n_{\frac{L_{w}}{2}-1}=\frac{2L_{\mathrm{p}}}{L_{w}}$, $a^{\langle i \rangle}_{j}\hspace{-1pt}:=a_{i+j\frac{L_{w}}{2}}$, $b^{\langle i \rangle}_{j}\hspace{-1pt}:=b_{i+j\frac{L_{w}}{2}}$, and $a_i$ and $b_i$ are the same as~(16), (17), \hspace{-0.071pt}and \hspace{-0.071pt}(18).
\hspace{-0.071pt}The \hspace{-0.071pt}matrices \hspace{-0.071pt}in \hspace{-0.071pt}the \hspace{-0.071pt}left \hspace{-0.071pt}side \hspace{-0.071pt}of \hspace{-0.071pt}(20) \hspace{-0.071pt}are \hspace{-0.071pt}\textit{periodic~\hspace{-0.071pt}tridiagonal matrices}, whose LU decompositions are also given in $\mathcal O(n)$ [37], and hence $\bm{x}_{\mathrm{p}}$ and the~final result $\bm{x}$ are also quickly obtained.

In particular, when $\mathrm{mod}_{\xi}(\frac{L_{w}}{2})=0$ for Type-I and Type-II,~\vspace{-0.25pt}or $\mathrm{mod}_{2\xi}(\frac{L_{w}}{2})\hspace{-0.22pt}=\hspace{-0.22pt}0$ for Type-III, we have \vspace{-0.25pt}$a^{\langle i \rangle}_{0}\hspace{-1.22pt}=\hspace{-0.22pt}a^{\langle i \rangle}_{1}\hspace{-1.22pt}=\hspace{-0.22pt}\cdots\hspace{-0.22pt}=\hspace{-0.22pt}a^{\langle i \rangle}_{n_{i}-1}\hspace{-1.22pt}=a_{i}$ and $b^{\langle i \rangle}_{0}\hspace{-1pt}=b^{\langle i \rangle}_{1}\hspace{-1pt}=\cdots=b^{\langle i \rangle}_{n_{i}-1}\hspace{-1pt}=b_{i}$, and hence the matrices in~the left side of (20) are \textit{symmetric circulant matrices}.
In such cases, the eigenvalues are \vspace{-1.5pt}$\lambda^{\langle i \rangle}_{q}\hspace{-1pt}=a_{i}+2b_{i}\cos(\frac{2q\pi}{n_{i}})\geq a_{i}-2|b_{i}|>0$ and the eigenvectors are
$\bm{u}_{q}=(1,e^{-\imath \frac{2q\pi}{n_{i}}},e^{-\imath \frac{4q\pi}{n_{i}}},\ldots,e^{-\imath  \frac{2(n_{i}-1)q\pi}{n_{i}}})^{\mathrm T}\in \mathbb C^{n_{i}}$
 ($q=0,1,\ldots,n_{i}-1$).
 Note that  Stankovi\'{c} defined the discrete-time window function  as $w[\tau]:=w(\tau \hspace{0.5pt}T_{\mathrm{s}})$ in \cite{Stankovic-16},\footnote{In \cite{Sondergaard-07}, the usual window function $w[\tau]=w(\tau \hspace{0.5pt}T_{\mathrm{s}})$ is called \textit{whole-point even (WPE)},
while $w[\tau]=w((\tau+\frac{1}{2})\hspace{0.25pt} T_{\mathrm{s}})$ is called \textit{half-point even~\vspace{-0.5pt}(HPE)}.} which results~in $|b_i|\hspace{-0.8pt}=\hspace{-0.8pt}\frac{a_i}{2}$ and $\lambda^{\langle i \rangle}_{q}\hspace{-1.7pt}=\hspace{-0.8pt}0$ for $i\hspace{-0.8pt}=\hspace{-0.8pt}\frac{L_w}{4}$ when $\xi\hspace{-0.8pt}=\hspace{-0.8pt}\frac{L_{w}}{2}$.
On the other~hand, we defined it as $w[\tau]:=w((\tau+\frac{1}{2})\hspace{0.25pt} T_{\mathrm{s}})$ in this paper, which guarantees $|b_i|<\frac{a_i}{2}$ and $\lambda^{\langle i \rangle}_{q}\hspace{-1pt}>0$ for all $i$. Hence, the inversion result~$\bm{x}_{\mathrm{p}}$ is also obtained by using FFT,
but we have to note that if $\xi\hspace{-0.8pt}=\hspace{-0.8pt}\frac{L_{w}}{2}$ and $L_{w}$ is relatively large, then $\mathcal S^{\mathrm{H}}_{\mathrm{p}}\circ \mathcal S_{\mathrm{p}}$ becomes ill-conditioned.\footnote{The condition number of $\mathcal S^{\mathrm{H}}_{\mathrm{p}}\circ \mathcal S_{\mathrm{p}}$ equals  $\max_{i,q}\{\lambda^{\langle i \rangle}_q\hspace{-0.5pt}\}/\min_{i,q}\{\lambda^{\langle i \rangle}_q\hspace{-0.5pt}\}$.}%

When \vspace{-0.25pt}$\mathrm{mod}_{\xi}(\frac{L_{w}}{2})=0$ and $\mathrm{mod}_{2\xi}(\frac{L_{w}}{2})\neq0$ for Type-III, we have $b^{\langle i \rangle}_{j}\hspace{-1pt}=(-1)^{j}b_{i}$ for all $i$ and $j$.
We can also compute the inversion result $\bm{x}_{\mathrm{p}}$ by using FFT, even in this case, with appropriate sign reversal process or appropriate multiplication process by $\imath$ depen-dently on~$\mathrm{mod}_{4}(n_{i})$ (see the actual program in \cite{program} for \vspace{-5.3pt}more~detail).

\subsection{Difference between the Two Inversions for FUSTFT\vspace{-1.2pt}}%}
Many papers explain the discrete STFT under the periodic condition as shown in Sect.~4.2.
In the cases of the discrete STFT and~FOSTFT, actually both $\mathcal S^{\mathrm{H}}\circ \mathcal S$ and $\mathcal S^{\mathrm{H}}_{\mathrm{p}}\circ \mathcal S_{\mathrm{p}}$ are diagonal matrices whose~diagonal components are periodic, and the two inversion results are~always
the same.
As a result, there is almost no problem even if we explain the discrete STFT and FOSTFT without the periodic condition.\footnote{In fact, in this paper, we derived the discrete STFT and FOSTFT including their inversions consistently, from the definitions of the continuous-time STFT and the discrete-time STFT, without assuming the periodic condition.%
}%

On the other hand, in the case of \vspace{-0.25pt}FUSTFT, these two inversions have different properties.
Define $\mathcal R:=\{\bm{X}\in\mathbb C^{\frac{L_{w}}{2}\times \left\lceil\frac{L_{x} + L_{w} - \xi}{\xi}\right\rceil}\,|$
$\exists\bm{x}\hspace{0.1858pt}\in\hspace{0.1858pt}\mathbb C^{L_x}\,\hspace{0.2858pt}\bm{X} \hspace{0.1858pt}=\hspace{0.1858pt}\mathcal S(\bm{x})\}$ \hspace{0.1858pt}as \hspace{0.1858pt}the \hspace{0.1858pt}\textit{range \hspace{0.1858pt}of \hspace{0.1858pt}FUSTFT} \hspace{0.1858pt}$\mathcal S$.
\hspace{0.1858pt}For \hspace{0.1858pt}a \hspace{0.1858pt}complex\pagebreak

\noindent spectrogram $\bm{X}\hspace{-0.7pt}\in\hspace{-0.7pt}\mathcal R$, both inversions can recover $\bm{x}$~s.t.~$\hspace{-0.1pt}\bm{X}\hspace{-0.7pt}=\hspace{-0.7pt}\mathcal S(\bm{x})$.\footnote{It is obvious for the case of $(\mathcal S^{\mathrm{H}}\circ\hspace{0.5pt} \mathcal S)^{-1}\circ \hspace{0.5pt}\mathcal S^{\mathrm{H}}$.
For $(\mathcal S^{\mathrm{H}}_{\mathrm p}\circ\hspace{0.5pt} \mathcal S_{\mathrm p})^{-1}\circ\hspace{0.5pt} \mathcal S^{\mathrm{H}}_{\mathrm p}$, the first $L_{x}$ components of $\bm{x}_{\mathrm p}$ become~$\bm{x}$, and the last $L_{\mathrm p}\hspace{-1pt}-\hspace{-0.5pt}L_{x}$ ones become~$\bm{0}$.}
For~$\bm{X}\not\in\mathcal R$, the standard inversion $(\mathcal S^{\mathrm{H}}\circ \mathcal S)^{-1}\circ \mathcal S^{\mathrm{H}}$~can recover the most consistent $\bm{x}$ with $\bm{X}$, while the inversion $(\mathcal S^{\mathrm{H}}_{\mathrm p}\circ\mathcal S_{\mathrm p})^{-1}\circ \mathcal S^{\mathrm{H}}_{\mathrm p}$ in the periodic condition can recover the most consistent $\bm{x}_{\mathrm{p}}$ with~$\bm{X}_{\mathrm{p}}$,  which means that the last $L_{\mathrm{p}}-L_x$ components of $\bm{x}_{\mathrm{p}}$ become non-zero.  
It might seem that the former inversion $(\mathcal S^{\mathrm{H}}\circ \mathcal S)^{-1}\circ \mathcal S^{\mathrm{H}}$~should always be used, but the latter inversion \vspace{-1pt}$(\mathcal S^{\mathrm{H}}_{\mathrm p}\circ\mathcal S_{\mathrm p})^{-1}\circ \mathcal S^{\mathrm{H}}_{\mathrm p}$ has a special property in the case of $\xi\hspace{-0.8pt}=\hspace{-0.8pt}\frac{L_w}{2}$.
Only when \vspace{0.4pt}$\xi\hspace{-0.8pt}=\hspace{-0.8pt}\frac{L_w}{2}$, $\mathcal S_{\mathrm p}$~\mbox{becomes} a~\textit{non-redundant} transform, and $\bm{x}_{\mathrm{p}}$~s.t.~$\bm{X}_{\mathrm{p}}=\mathcal S_{\mathrm p}(\bm{x}_{\mathrm{p}})$ can be always recovered by the latter inversion.
As a result, for any complex~spectrogram $\bm{X}\not\in\mathcal R$, a discrete-time signal $\bm{x}$ that guarantees~the~\textit{perfect consistency other than the first and last frames} is always  recovered.\footnote{This is the same property that the inversions for MDCT and DWT have.}%

\begin{table}[t]
\centering
\begin{minipage}{1\hsize}%
\centering
\vspace{-6.4pt}
\caption{Inversion results $\hat{\bm{x}}$ from $\bm{X}\in\mathcal R$ which is inside the range.}\vspace{-0.5pt}%
{\scriptsize \begin{tabular}{|c|c|c|c|c|c|} \hline
$L_w$ &$\xi$& \hspace{-3.5pt}Inverse\hspace{-3.5pt}\mbox{}&$\frac{\lVert \bm{x}-\hat{\bm{x}}\rVert_2}{\lVert \bm{x}\rVert_2}$ & \hspace{-2pt}$\frac{\lVert \bm{X}-\mathcal S(\hat{\bm{x}})\rVert_{\mathrm F}}{\lVert \bm{X}\rVert_{\mathrm F}}\hspace{-2pt}$&  $\begin{aligned}\mbox{}\\[-10pt]\hspace{-4pt}\tfrac{\lVert \bm{X}-\mathcal S(\hat{\bm{x}})\rVert^{\mathrm{int}}_{\mathrm F}}{\lVert \bm{X}\rVert_{\mathrm F}^{\mathrm{int}}}\hspace{-4pt}\end{aligned}$\\[5.25pt]\hline%
        \multirow{4}{*}{\hspace{-2pt}$\begin{aligned}\mbox{}\\[-4.5pt]2{,}048\end{aligned}\hspace{-2pt}$} &\multirow{2}{*}{\hspace{-2pt}$\frac{L_w}{2}\hspace{-2pt}$} &
        $\begin{aligned}\mbox{}\\[-8.5pt]
        \mathcal S\\[-10pt]
        \mbox{}
        \end{aligned}$&\hspace{-2pt}$5.0\times 10^{-14}\hspace{-3pt}$&\hspace{-2pt}$8.3\times 10^{-16}\hspace{-3pt}$ &\hspace{-2pt}$8.3\times 10^{-16}\hspace{-3pt}$\\[-1pt]\cline{3-6}%[5pt]\hline
      &  &$\begin{aligned}\mbox{}\\[-9pt]
        \mathcal S_{\mathrm p}
        \\[-9pt]
        \mbox{}\end{aligned}$&\hspace{-2pt}\textcolor{blue}{$1.0\times 10^{-13}\hspace{-3pt}$}&\hspace{-2pt}\textcolor{blue}{$3.8\times 10^{-15}\hspace{-3pt}$} &\hspace{-2pt}$9.6\times 10^{-16}\hspace{-3pt}$\\[-1pt]\cline{2-6}%[5pt]\hline
%    &\multirow{2}{*}{$\frac{L_w}{4}$} &$\mathcal S$&$27.97$&$29.37$ &$\bm{29.73}$\\\cline{3-6}
%  &  &$\mathcal S_{\mathrm{p}}$&$27.97$&$29.37$ &$\bm{29.73}$\\\cline{2-6}
  &\multirow{2}{*}{\hspace{-3pt}$\frac{L_w}{8}\hspace{-3pt}$} &$\begin{aligned}\mbox{}\\[-8pt]
        \mathcal S\\[-10pt]
        \mbox{}
        \end{aligned}$&\hspace{-2pt}$6.5\times 10^{-16}\hspace{-3pt}$&\hspace{-2pt}$6.7\times 10^{-16}\hspace{-3pt}$ &\hspace{-2pt}$6.7\times 10^{-16}\hspace{-3pt}$\\[-1.25pt]\cline{3-6}%[5pt]\hline
  &  &$\begin{aligned}\mbox{}\\[-9pt]
        \mathcal S_{\mathrm p}
        \\[-9pt]
        \mbox{}\end{aligned}$&\hspace{-2pt}$4.7\times 10^{-16}\hspace{-3pt}$&\hspace{-2pt}$5.1\times 10^{-16}\hspace{-3pt}$ &\hspace{-2pt}$5.1\times 10^{-16}\hspace{-3pt}$\\[-1pt]\hline
%  \multirow{6}{*}{$4{,}096$} &\multirow{2}{*}{$\frac{L_w}{2}$} &$\mathcal S$&$27.97$&$29.37$ &$\bm{29.73}$\\\cline{3-6}%[5pt]\hline
%      &  &$\mathcal S_{\mathrm{p}}$&$27.97$&$29.37$ &$\bm{29.73}$\\\cline{2-6}
%    &\multirow{2}{*}{$\frac{L_w}{4}$} &$\mathcal S$&$27.97$&$29.37$ &$\bm{29.73}$\\\cline{3-6}
%  &  &$\mathcal S_{\mathrm{p}}$&$27.97$&$29.37$ &$\bm{29.73}$\\\cline{2-6}
%  &\multirow{2}{*}{$\frac{L_w}{8}$} &$\mathcal S$&$27.97$&$29.37$ &$\bm{29.73}$\\\cline{3-6}
%  &  &$\mathcal S_{\mathrm{p}}$&$27.97$&$29.37$ &$\bm{29.73}$\\\hline
  \multirow{4}{*}{\hspace{-2pt}$\begin{aligned}\mbox{}\\[-4.5pt]8{,}192\end{aligned}\hspace{-2pt}$} &\multirow{2}{*}{\hspace{-2pt}$\frac{L_w}{2}\hspace{-2pt}$} &
        $\begin{aligned}\mbox{}\\[-8.5pt]
        \mathcal S\\[-10pt]
        \mbox{}
        \end{aligned}$&\hspace{-2pt}$1.3\times 10^{-14}\hspace{-3pt}$&\hspace{-2pt}$5.7\times 10^{-16}\hspace{-3pt}$ &\hspace{-2pt}$5.7\times 10^{-16}\hspace{-3pt}$\\[-1pt]\cline{3-6}%[5pt]\hline
      &  &$\begin{aligned}\mbox{}\\[-9pt]
        \mathcal S_{\mathrm p}
        \\[-9pt]
        \mbox{}\end{aligned}$&\hspace{-2pt}\textcolor{blue}{$1.8\times 10^{-12}\hspace{-3pt}$}&\hspace{-2pt}\textcolor{blue}{$1.1\times 10^{-13}\hspace{-3pt}$} &\hspace{-2pt}$9.8\times 10^{-16}\hspace{-3pt}$\\[-1pt]\cline{2-6}%[5pt]\hline
%    &\multirow{2}{*}{$\frac{L_w}{4}$} &$\mathcal S$&$27.97$&$29.37$ &$\bm{29.73}$\\\cline{3-6}
%  &  &$\mathcal S_{\mathrm{p}}$&$27.97$&$29.37$ &$\bm{29.73}$\\\cline{2-6}
  &\multirow{2}{*}{\hspace{-2pt}$\frac{L_w}{8}\hspace{-2pt}$} &$\begin{aligned}\mbox{}\\[-8pt]
        \mathcal S\\[-10pt]
        \mbox{}
        \end{aligned}$&\hspace{-2pt}$4.4\times 10^{-16}\hspace{-3pt}$&\hspace{-2pt}$5.1\times 10^{-16}\hspace{-3pt}$ &\hspace{-2pt}$5.1\times 10^{-16}\hspace{-3pt}$\\[-1.25pt]\cline{3-6}%[5pt]\hline
  &  &$\begin{aligned}\mbox{}\\[-9pt]
        \mathcal S_{\mathrm p}
        \\[-10pt]
        \mbox{}\end{aligned}$&\hspace{-2pt}$
        \begin{aligned}\mbox{}\\[-9.75pt]4.4\times 10^{-16}
        \end{aligned}\hspace{-3pt}$&\hspace{-2pt}$\begin{aligned}\mbox{}\\[-9.75pt]5.0\times 10^{-16}\end{aligned}\hspace{-3pt}$ &\hspace{-2pt}$\begin{aligned}\mbox{}\\[-9.75pt]5.0\times 10^{-16}\end{aligned}\hspace{-3pt}$\\\hline
	\end{tabular}}%
	 \end{minipage}\\%\\% %\ \,%
	 \begin{minipage}{1\hsize}%
	 \centering
	 \vspace{-8.9pt}
	% \vspace{-10pt}
	\caption{Inversion results $\hat{\bm{x}}$ from $\widehat{\bm{X}}\not\in\mathcal R$ which is outside the range.}\vspace{-0.5pt}
{\scriptsize\begin{tabular}{|c|c|c|c|c|c|} \hline
$L_w$ &$\xi$& \hspace{-3.5pt}Inverse\hspace{-3.5pt}\mbox{}&$\frac{\lVert \bm{x}-\hat{\bm{x}}\rVert_2}{\lVert \bm{x}\rVert_2}$ & \hspace{-2pt}$\frac{\lVert \widehat{\bm{X}}-\mathcal S(\hat{\bm{x}})\rVert_{\mathrm F}}{\lVert \widehat{\bm{X}}\rVert_{\mathrm F}}\hspace{-2pt}$&  $\begin{aligned}\mbox{}\\[-10pt]\hspace{-4pt}\tfrac{\lVert \widehat{\bm{X}}-\mathcal S(\hat{\bm{x}})\rVert^{\mathrm{int}}_{\mathrm F}}{\lVert \widehat{\bm{X}}\rVert_{\mathrm F}^{\mathrm{int}}}\hspace{-4pt}\end{aligned}$\\[5.25pt]\hline%
        \multirow{4}{*}{\hspace{-2pt}$\begin{aligned}\mbox{}\\[-4.5pt]2{,}048\end{aligned}\hspace{-2pt}$} &\multirow{2}{*}{\hspace{-2pt}$\frac{L_w}{2}\hspace{-2pt}$} &
         $\begin{aligned}\mbox{}\\[-8.5pt]
        \mathcal S\\[-10pt]
        \mbox{}
        \end{aligned}$&\hspace{-2pt}$2.9\times 10^{-2}\hspace{-3pt}$&\hspace{-2pt}$6.0\times 10^{-4}\hspace{-3pt}$ &\hspace{-2pt}$2.5\times 10^{-4}\hspace{-3pt}$\\[-1pt]\cline{3-6}%[5pt]\hline
      &  &$\begin{aligned}\mbox{}\\[-9pt]
        \mathcal S_{\mathrm p}
        \\[-9pt]
        \mbox{}\end{aligned}$&\hspace{-2pt}$3.5\times 10^{-2}\hspace{-3pt}$&\hspace{-2pt}$1.2\times 10^{-3}\hspace{-3pt}$ &\hspace{-2pt}\textcolor{red}{$1.7\times 10^{-15}\hspace{-3pt}$}\\[-1pt]\cline{2-6}%[5pt]\hline
  &\multirow{2}{*}{\hspace{-2pt}$\frac{L_w}{8}\hspace{-2pt}$} &$\begin{aligned}\mbox{}\\[-8pt]
        \mathcal S\\[-10pt]
        \mbox{}
        \end{aligned}$&\hspace{-2pt}$8.0\times 10^{-3}\hspace{-3pt}$&\hspace{-2pt}$1.3\times 10^{-2}\hspace{-3pt}$ &\hspace{-2pt}$1.3\times 10^{-2}\hspace{-3pt}$\\[-1.25pt]\cline{3-6}%[5pt]\hline
  &  &$\begin{aligned}\mbox{}\\[-9pt]
        \mathcal S_{\mathrm p}
        \\[-9pt]
        \mbox{}\end{aligned}$&\hspace{-2pt}$8.0\times 10^{-3}\hspace{-3pt}$&\hspace{-2pt}$1.3\times 10^{-2}\hspace{-3pt}$ &\hspace{-2pt}$1.3\times 10^{-2}\hspace{-3pt}$\\[-1pt]\hline
  \multirow{4}{*}{\hspace{-2pt}$\begin{aligned}\mbox{}\\[-4.5pt]8{,}192\end{aligned}\hspace{-2pt}$} &\multirow{2}{*}{\hspace{-2pt}$\frac{L_w}{2}\hspace{-2pt}$} &
         $\begin{aligned}\mbox{}\\[-8.5pt]
        \mathcal S\\[-10pt]
        \mbox{}
        \end{aligned}$&\hspace{-0.225pt}$2.7\times 10^{-2}\hspace{-1.225pt}$&\hspace{-0.225pt}$1.6\times 10^{-3}\hspace{-1.225pt}$ &\hspace{-2pt}$4.2\times 10^{-4}\hspace{-3pt}$\\[-1pt]\cline{3-6}%[5pt]\hline
      &  &$\begin{aligned}\mbox{}\\[-9pt]
        \mathcal S_{\mathrm p}
        \\[-9pt]
        \mbox{}\end{aligned}$&\hspace{-2pt}$6.9\times 10^{-2}\hspace{-3pt}$&\hspace{-2pt}$4.5\times 10^{-3}\hspace{-3pt}$ &\hspace{-2pt}\textcolor{red}{$1.1\times 10^{-14}\hspace{-3pt}$}\\[-1pt]\cline{2-6}%[5pt]\hline
  &\multirow{2}{*}{\hspace{-3pt}$\frac{L_w}{8}\hspace{-3pt}$} &$\begin{aligned}\mbox{}\\[-8pt]
        \mathcal S\\[-10pt]
        \mbox{}
        \end{aligned}$&\hspace{-2pt}$8.0\times 10^{-3}\hspace{-3pt}$&\hspace{-2pt}$1.4\times 10^{-2}\hspace{-3pt}$ &\hspace{-2pt}$1.3\times 10^{-2}\hspace{-3pt}$\\[-1.25pt]\cline{3-6}%[5pt]\hline
  &  &$\begin{aligned}\mbox{}\\[-9pt]
        \mathcal S_{\mathrm p}
        \\[-10pt]
        \mbox{}\end{aligned}$&\hspace{-2pt}$\begin{aligned}\mbox{}\\[-9.75pt]8.0\times 10^{-3}\end{aligned}\hspace{-3pt}$&\hspace{-2pt}$\begin{aligned}\mbox{}\\[-9.75pt]1.4\times 10^{-2}\end{aligned}\hspace{-3pt}$ &\hspace{-2pt}$\begin{aligned}\mbox{}\\[-9.75pt]1.3\times 10^{-2}\end{aligned}\hspace{-3pt}$\\\hline
	\end{tabular}}%
		 \end{minipage}%\\[6.75pt]%\ \ \,%
\vspace{-12.3pt}%
\end{table}

We confirm the properties of the inversions in Sects.~4.1 and~4.2 for Type-II FUSTFT\@. 
A sound signal $\bm{x}$ of $15$ seconds is transformed into a~complex spectrogram \vspace{-1pt}$\bm{X}\hspace{-0.3pt}\in\hspace{-0.3pt}\mathcal R$,\footnote{We used a male voice, that counts numbers, of $f_{\mathrm{s}}=44{,}100$ [Hz] in~\cite{mat} and  the normalized Hann window \vspace{-0.5pt}$w[\tau]:=\frac{1}{2\sqrt{L_w}}(1 - \cos(\frac{2}{L_w}(\tau+\frac{1}{2})\pi))$.} 
and we create a noisy version $\widehat{\bm{X}}\not\in\mathcal R$ by adding complex white Gaussian noise of variance \vspace{-1pt}$10^{-6}$ to $\bm{X}$.
The performance of the inversion results $\hat{\bm{x}}$ from~$\bm{X}$ and $\widehat{\bm{X}}$ are summarized in Tables 1 and 2,\footnote{In Tables 1 and 2, we call $\lVert\cdot\rVert_{\mathrm F}^{\mathrm{int}}$ the \textit{interior} Frobenius norm that
ignores components in the first $\bigl\lceil\frac{L_w-\xi}{\xi}\bigr\rceil$ and last $\bigl\lceil\bigl\lceil\frac{L_{x} + L_{w} - \xi}{\xi}\bigr\rceil  - \frac{L_{x}}{\xi}\bigr\rceil$ time frames.} respectively.
From these tables, we can confirm that both inversions work correctly.
In particular,~when $\xi=\frac{L_w}{2}$, the inversion results by $(\mathcal S^{\mathrm{H}}_{\mathrm p}\circ\mathcal S_{\mathrm p})^{-1}\circ \mathcal S^{\mathrm{H}}_{\mathrm p}$  demonstrated~the

\noindent  slight numerical instability in blue letters of Table 1 and the perfect consistency \vspace{-7pt} \hspace{-0.223pt}other \hspace{-0.223pt}than \hspace{-0.223pt}the \hspace{-0.223pt}first \hspace{-0.223pt}and \hspace{-0.223pt}last \hspace{-0.223pt}frames \hspace{-0.223pt}in \hspace{-0.223pt}red \hspace{-0.223pt}letters \hspace{-0.223pt}of \hspace{-0.223pt}Table~\hspace{-0.223pt}2.

\section{Conclusion\vspace{-5pt}}
This paper proposed FUSTFT and its two inversions. FUSTFT gives efficient spectrogram matrices by
computing only half the frequency components of the discrete STFT\@.
By using FUSTFT, it is~expected that window functions of relatively large $L_w$ such as the Kaiser window and the truncated Gaussian window will be easier to use.
Since we can arbitrarily modify magnitudes and phases other than the first and last frames when $\xi=\frac{L_w}{2}$, further development of spectrogram-based \hspace{-0.137pt}techniques \hspace{-0.137pt}is \hspace{-0.137pt}also \hspace{-0.137pt}expected.
\hspace{-0.137pt}There \hspace{-0.137pt}is \hspace{-0.137pt}also \hspace{-0.137pt}a \hspace{-0.137pt}possibility \hspace{-0.137pt}that~\hspace{-0.137pt}flexible transforms such as Type-III FUSTFT will find new applications.\pagebreak%

\pagebreak

\end{document}